\documentclass[a4paper,11pt]{article}
\usepackage{jinstpub}
\usepackage{subcaption}
\usepackage{comment}
\usepackage{mathtools}
\usepackage{url}
\usepackage[per-mode=symbol,separate-uncertainty=true, exponent-product=\cdot,multi-part-units=brackets,product-units=power]{siunitx}

\title{Calibration of 122 SensL MicroFJ-60035 SiPMs and the reduction of optical crosstalk due to coupled light guides}

\author[a,1]{F.~Rehbein,\note{Corresponding author.}}
\author[a,1]{T.~Bretz,}
\author[b]{R.~Alfaro,}
\author[a]{J.~Audehm,}
\author[c,1]{A.~Biland,}
\author[a]{G.~Do,}
\author[d]{M.~M.~Gonz\'alez,}
\author[d]{Y.~F.~P\'erez-Araujo,}
\author[e]{M.~Schaufel,}
\author[b]{J.~Serna-Franco,}
\author[f]{I.~Torres}

\affiliation[a]{Physics Institute 3A, RWTH Aachen University, Sommerfeldstra\ss e 16, 52074 Aachen, Germany.}
\affiliation[b]{Instituto de F\'isica, Universidad Nacional Aut\'onoma de M\'exico, Ciudad de M\'exico, Mexico.}
\affiliation[c]{ETH Zurich, Inst. for Particle Physics and Astrophysics, Otto-Stern-Weg 5, 8093 Zurich, Switzerland.}
\affiliation[d]{Instituto de Astronom\'ia, Universidad Nacional Aut\'onoma de M\'exico, Ciudad de M\'exico, Mexico.}
\affiliation[e]{Physics Institute 3B, RWTH Aachen University, Sommerfeldstra\ss e 16, 52074 Aachen, Germany.}
\affiliation[f]{Instituto Nacional de Astrof\'isica, \'Optica y Electr\'onica, Puebla, Mexico.}

\emailAdd{florian.rehbein@rwth-aachen.de}

\abstract{
The excellent production quality of recent generations of Silicon Photomultipliers (SiPMs) allows for operation without individual calibration of the breakdown voltage. Measurements of the crosstalk probability and the relative gain of 122~SiPMs of type SensL MicroFJ-60035-TSV are presented. Semi-conductor photo sensors have replaced photo multiplier tubes in numerous applications featuring single-photon resolution, insensitivity to magnetic fields, higher robustness and enhanced photo detection efficiency at lower operation voltage and lower costs. Light guides are used to increase the comparably small photo sensitive area of SiPMs. Their optical coupling changes the surface conditions of the sensor and influences the probability for secondary photons to leave the sensor without inducing secondary breakdowns. This study compares properties of sensors that are optically coupled to light guides with bare sensors, operated at nominal bias voltage. It demonstrates, that the optical coupling to a light guide significantly reduces the crosstalk probability of the measured sensors.
}


\begin{document} 
\maketitle
\flushbottom

\section{Introduction}
\label{sec::intro}
In many applications, the ability to detect very faint light flashes requires photo sensors with a high photo detection efficiency and a high level of robustness. Silicon Photomultipliers (SiPMs) provide durability, low operation voltage, and robustness against bright light exposure. For instance, in imaging air-Cherenkov astronomy, SiPMs were initially applied in the First G-APD Cherenkov Telescope (FACT~\cite{Anderhub_2013}) as a replacement for classical photo multiplier tubes. The telescope is successfully operated now for almost ten years~\cite{2019ICRC...36..665D}. Its ability to operate even under bright background illumination, such as moonlight~\cite{2013ICRC...33.1132K}, provides an increased duty cycle~\cite{2019ICRC...36..665D} of almost \SI{30}{\percent}. The robustness and stability of the sensors allow for consistent unbiased monitoring of astronomical sources enabling unprecedented long-term studies (e.\ g.~\cite{2019ICRC...36..630B}).\\

Significant improvement of the photo detection efficiency of SiPMs during the recent decades have facilitated the construction of compact telescopes with an aperture diameter of only \SI{50}{\centi\meter}~\cite{Bretz_2018}. They are also the first imaging air-Cherenkov telescopes utilizing refractive optics based on a Fresnel lens in operation. Based on this design, similar telescopes were constructed to achieve hybrid measurements together with extensive air shower arrays. Hybrid measurements allow resolving the ambiguity on shower age, intrinsic to air shower arrays, and on shower distance, intrinsic to telescopes, enabling for inexpensive low-resolution optics. Measurements are ongoing with the IceAct telescope at the IceCube neutrino observatory at the South Pole, for example, for the improvement of veto capabilities~\cite{2020APh...11702417R}, and together with the High Altitude-Water Cherenkov Observatory (HAWC~\cite{2014BrJPh..44..571M}) in Mexico for calibration and increased sensitivity~\cite{ICRC}. 
Compared with the original design, these telescopes implement further improvements, including solid light guides with an increased collection efficiency and SiPMs with further enhanced photo detection efficiency. This study discusses results obtained with two telescopes at the HAWC site: named the HAWC's Eye telescopes.\\

The following paragraphs only summarize the properties, that are relevant for the context of this paper, and do not aim to present a complete description of the functioning and characteristics of SiPMs. A detailed review of Silicon Photomultipliers is given in~\cite{klanner2019}. They combine multiple Geiger-mode avalanche photo diodes (G-APD) into one sensor. Geiger-mode avalanche photo diodes are also commonly known as single-photon avalanche diodes (SPAD). 
In the following, the individual diodes will be referred to as cells. The diodes are operated in Geiger-mode, i.\ e.\ above a characteristic voltage, the so-called breakdown voltage. Above this threshold, impinging photons are generating electron-hole pairs, which may induce a self-sustaining particle cascade. The initiated current causes a voltage drop at the built-in quenching resistor. When the applied voltage falls below the breakdown voltage, the avalanche is stopped. The released charge is independent of the number of initial electron-hole pairs, and depends only on physical properties of the diode and the difference between the applied bias voltage and the breakdown voltage. The charge released by one discharge is commonly referred to as photon equivalent (p.\ e.). The voltage difference is known as overvoltage. The released charge is also independent of the energy and angle of the incident photon. Random discharges triggered by thermal excitation are known as dark counts.\\

Secondary discharges that are induced by trapped charge carriers are called afterpulses.
Their probability drops exponentially after the initial breakdown. As the time constant of this exponential decay is typically in the same order as the recharge time of the cell and the released charge is intrinsically limited by the charge state of the cell, contributions from afterpulses are highly suppressed during the early recharge phase. For more details see~\cite{klanner2019}. In the present study, the analysis is based on charge integration over only a small time-window. Therefore, the effect of afterpulses can be neglected. Furthermore, the total afterpulse probability of recent sensors is less than a few percent\footnote{$<5\,\%$ for SensL MicroFJ-60030-TSV at \SI{5}{\volt} overvoltage~\cite{OnSemi}}.\\

The arrangement of hundreds to tens of thousands of cells in one sensor may introduce additional noise features into the measured signal, e.\ g.\ optical crosstalk. Optical crosstalk is induced by photons emitted during recombination triggering secondary breakdowns in neighboring cells. This alters the amplitude of the measured signal. For high multiplicities, it contributes a statistical increase. Usually, a low crosstalk probability can only be achieved by a selection of a device type specifically built to achieve low crosstalk probability. Modern SiPMs comprise thin optical trenches that surround the active area of the cells to reduce direct crosstalk between neighboring cells~\cite{6044204}. The crosstalk probability refers to the fraction of signals with at least one secondary breakdown out of all signals. A detailed description of different photon paths relevant to the crosstalk process is given in~\cite{Rech:08}. This paper targets particularly those photons reflected at the protective window of the sensor. A measurement which spatially resolves crosstalk photons exiting the protective window of a SiPM is presented in~\cite{2018EPJC...78..971E}. 
A special case is the so-called delayed crosstalk, see~\cite{klanner2019} for more details. Less probable than afterpulses in the same cell, the effect of delayed crosstalk photons on a small integration window can be neglected as well.\\

Delayed crosstalk that comes early enough to be included in the integration window just contributes to the relevant crosstalk of the signal. Delayed crosstalk that comes with a delay longer than the integration window does not contribute to the signal and can thus be ignored.

The following study demonstrates that a well-designed light guide which is optically coupled to the protective window decreases crosstalk probability. This is achieved by coupling a transparent material with a comparable refractive index to the protective window, suppressing total reflection. Due to the special shape of the coupled light guide, these photons can be directed out of its entrance aperture, thus avoiding triggers in neighboring cells.\\

\section{Experimental Setup} \label{section:hardware}

\begin{figure}[ht]
\centering
\includegraphics[width=0.7\textwidth]{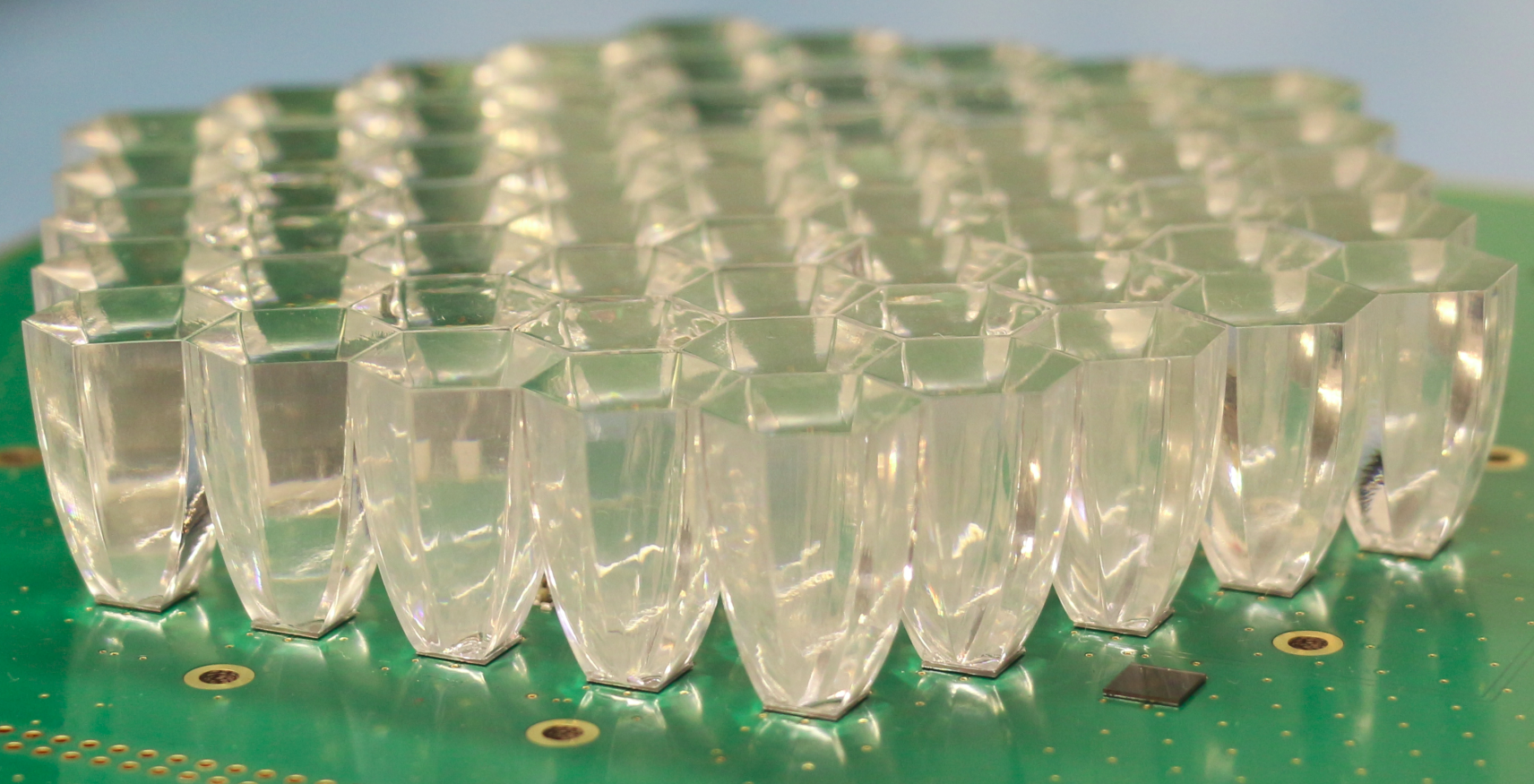}
\caption{Picture of the camera of the HAWC's Eye telescope. 61~SiPMs are optically coupled to solid light guides. One of the three blind pixels can be seen on the bottom right of the picture.}
\label{fig:camera}
\end{figure}

For the presented measurements, two telescopes of the HAWC's Eye type are utilized. They are located at the HAWC site in Mexico and can be operated remotely. The camera of each telescope features 64~sensors, out of which 61~sensors are optically coupled to solid light guides to increase their light collection area. The remaining three SiPMs ("blind pixels") are used for calibration purposes. Figure~\ref{fig:camera} shows a picture of the camera of one of the telescopes prior to its assembly.  Due to transport damage, one of the two cameras misses one sensor and three light guides. Moreover, one of its light guides remains loosely on the respective SiPM. The utilized data acquisition is identical to the system installed in the FACT telescope. A detailed description of the readout circuit can be found in~\cite{Anderhub_2013}.\\

\subsection*{Sensors and Bias Power Supply}

The utilized sensors are of type SensL\footnote{now OnSemi} MicroFJ-60035-TSV~\cite{OnSemi} which are operated at an overvoltage of \SI{5}{\volt}. The relevant specifications of the sensor are summarized in Table~\ref{tab:sipm_specifications}. Each sensor has an active area of $6.07\,\times\,6.07$\,\si{\square\milli\meter} and consists of \num{22292}~cells. They are composed of a thin layer of silicon with a thickness of \SI{0.09}{\milli\meter} and a protective window of \SI{0.37}{\milli\meter} thickness. The protective window is made of a glass substrate with a refractive index of 1.53 at a reference wavelength of \SI{436}{\nano\meter}. The sensors achieve a peak photo detection efficiency of \SI{50}{\percent} at a wavelength of \SI{420}{\nano\meter} if operated at \SI{6}{\volt} overvoltage above their nominal breakdown voltage of \SI{24.5}{\volt}. The data sheet defines the minimum and maximum breakdown voltage as \SI{24.2}{\volt} and \SI{24.7}{\volt} at a temperature of \SI{21}{\degreeCelsius}. This range corresponds to a maximum gain variation of \SI{\pm\,5}{\percent} at \SI{5}{\volt} overvoltage. It is expected that the typical sensor-to-sensor variation of breakdown voltages is smaller. The bias voltage is supplied from a power supply based on the design discussed in~\cite{ICRC_2017_schumacher}. 
It adjusts the applied voltage of each sensor individually to compensate for a change of breakdown voltage with temperature. The temperature sensors provide an absolute precision of \SI{1}{K} without additional calibration. In total, 64 temperature sensors are located on the backside of the printed circuit board opposite the SiPMs. It is implied that the temperature sensors and the SiPMs are in thermal equilibrium, as the typical dark current of only a few \si{\micro\ampere} corresponds to a negligible heating capacity and no significant temperature gradient has been measured between individual sensors. In total, this system is expected to provide a precision of the temperature measurement better than \SI{2}{\kelvin} corresponding to a systematic relative deviation in overvoltage of less than \SI{1}{\percent} for the given temperature coefficient of \SI{21.5}{\milli\volt\per\kelvin} at \SI{5}{\volt} overvoltage. After calibration and without load, each channel of the bias power supply provides a voltage with a precision of typically \SI{2}{\milli\volt}, but not higher than \SI{10}{\mV}. Compared to the overvoltage, the precision of the power supply can therefore be neglected. The uncertainty of the gain is therefore limited by the uncertainty of the breakdown voltage and yields a total systematic relative uncertainty of the gain of \SI{5}{\percent}. When accounting for the applied overvoltage and the sensor temperature of \SI{28}{\degreeCelsius} (see Chapter~\ref{sec:method}), the dark count rate is expected to be in the order of \SI{200}{\kilo\hertz\per\square\mm}.
As shown in the following, dark count measurements allow for determining sensor properties such as the crosstalk probability and the gain. This helps to characterize and calibrate the SiPMs.\\ 

\begin{table}[]
    \centering
    \caption{Specifications of the SensL MicroFJ-60035-TSV sensors according to the data sheet provided by the manufacturer~\cite{OnSemi}. The performance parameters are given for an overvoltage of \SI{2.5}{\volt} (\SI{6}{\volt}) at a temperature of \SI{21}{\degreeCelsius}.}
    \begin{tabular}{|l|c|}
    \hline
         \textbf{Parameter} &  \textbf{Value}\\
         \hline
         Active Area [\si{\milli\meter\squared}] & $6.07 \times 6.07$\\
         Number of cells & \num{22292}\\
         Fill Factor & \SI{75}{\percent}\\
         Breakdown Voltage $V_{br}$ [\si{\volt}] & $24.45 \pm 0.25$ \\
         Temperature coefficient of $V_{br}$ [\si{\milli\volt\per\degreeCelsius}]& 21.5 \\
         Peak sensitivity wavelength $\lambda_p$ [\si{\nano\meter}] & 420\\
         PDE at  $\lambda_p$ &  $38 \, \%$ ($50 \, \%$)\\
         Recharge Time Constant [\si{\nano\second}] & 50\\
         Gain & $2.9 \times 10^6$ ($6.3 \times 10^6$)\\
         Dark Count Rate [\si{\kilo\hertz\per\milli\meter\squared}] & 50 (150)\\
         Dark Current (typical) [\si{\micro\ampere}] & 0.9 (7.5)\\
         Crosstalk probability& $8 \, \%$ ($25 \, \%$)\\
         Afterpulsing probability& $0.75 \, \%$ ($5.0 \, \%$)\\
          \hline
    \end{tabular}
    \label{tab:sipm_specifications}
\end{table}

\subsection*{Light Guides}

Light guides are often used in SiPM applications to increase their small photosensitive area. They are particularly common in SiPM cameras to decrease dead space at reduced total costs. In the camera of the FACT telescopes, solid light guides are successfully applied since 2011~\cite{Anderhub_2013, 1352692}.

The required acceptance angle of the applied light guides is usually defined by the numerical aperture of the imaging optics and by the requirements on the attenuation of stray light. Their exit aperture is given by the SiPM surface area. These constraints and the acceptance angle of the SiPM inherently limit their maximum concentration factor, since according to Liouville's theorem, the phase-space distribution function is constant along the trajectories of the system. As the refractive index of the material changes the effective path length of the trajectories, the maximum concentration factor is proportional to the refractive index $n$ squared, i.\ e.\ 
\begin{equation}
    \frac{A}{a} \propto {n}^2
\end{equation}
with the entrance aperture $A$ and the exit aperture $a$. The acceptance angle of the sensor is always limited by the critical angle for total reflection at the silicon surface.\\

The light guides of the HAWC's Eye camera are made of Plexiglas with a refractive index which is comparable to the one of the protective window of the sensors. The refractive index of the Plexiglas at \SI{436}{\nano\meter} is 1.50~\cite{Beadie:15} and for the sensor 1.53~\cite{OnSemi}. The shape of the light guides was designed to support an opening angle of \ang{33.7} corresponding to the view of an edge pixel at \SI{60}{\milli\meter} distance from the camera center and a numerical aperture  $f/D=502\,\text{mm}\,/\,550\,\text{mm}=0.9$ of the optical system \cite{Koschinsky_Master}. This leads to a compression factor of 5.5. With their refractive index of $n\approx1.5$, they achieve a concentration factor about as twice as large when compared to a corresponding hollow light guide, see also a more detailed discussion in~\cite{2013APh....45...44B}. The shape of the light guides is characterized by a hexagonal entrance window with a radius \SI{7.4}{\milli\meter} and a square output window. The length of the light guide is \SI{23.5}{\milli\meter}. The hexagonal aperture has been chosen to minimizing dead spaces in the camera, the square output window matches the sensor size. Four edges of the hexagon are connected to the four edges of the quadratic base. The two parallel sides form a Winston surface each, i.\ e. a tilted parabola with its focal point at the opposite side of the square surface~\cite{Winston_2004}. The remaining two edges are connected to the center of the two remaining sides of the square and also follow the Winston principle. The four remaining side walls are constructed with straight lines connecting the edges at any height. All light-guides have been manufactured with a CNC milling machine and manually polished.  Each light guide is optically coupled to its sensor with a very thin layer of optical glue of similar refractive index. All light guides have been glued manually allotting the glue with a dispensing pipette. The light guides have been placed carefully with the help of a fineplacer. None of the joints show impurities or other flaws. The following study shows comparable results of all those pixels coupled to a light guide and therefore no indication for relevant fluctuations of the coupling quality. \\

If the overall performance of solid and hollow light guides is evaluated for the case of an optimized concentration factor, also reflection and absorption have to be considered: Total reflection outperforms coated or polished surfaces. Internal absorption for Plexiglas is in the order of \SI{1}{\percent\per\centi\meter} and can be neglected\footnote{Note that absorption in data sheets usually includes the effect of Fresnel reflection at both surfaces.}. If light guides are covered by a protective window which is optically not coupled, an additional Fresnel loss of about \SI{4}{\percent} has to be taken into account at each window-air interface. This is always the case in the application of hollow light guides.

Optimized light guides are designed such that the maximum acceptance angle of the silicon is exhausted by incoming rays. That means that for each outgoing ray, at least one light path exists that leads the photon out of the guide. Consequently, optical coupling to a light guide leads to a reduced probability for these photons to be reflected back into the sensor. Contrary, in case of the application of non-optimized guides, photons might be internally reflected back to the sensor and trigger random cells instead of direct neighbors. These cells have a higher probability to be fully charged, as direct neighbors might have been discharged by other crosstalk photons already. For the first crosstalk photon, this is not the case, as neighbors are generally charged. 
Therefore these reflected photons increase the probability for higher multiplicities which means that non-optimized light guides can even increase the crosstalk probability.

\section{Method} \label{sec:method}

To measure the crosstalk probability, the signal induced by dark counts can be utilized. The excellent single-p.\ e. resolution of SiPMs allows distinguishing between 
signals generated by a different number of discharges. To obtain such a dark count spectrum, the integrated charge of random pulses induced by thermal excitation are extracted from the signal traces and filled into a histogram.

\begin{figure}[ht]
\centering
\includegraphics[width=\textwidth]{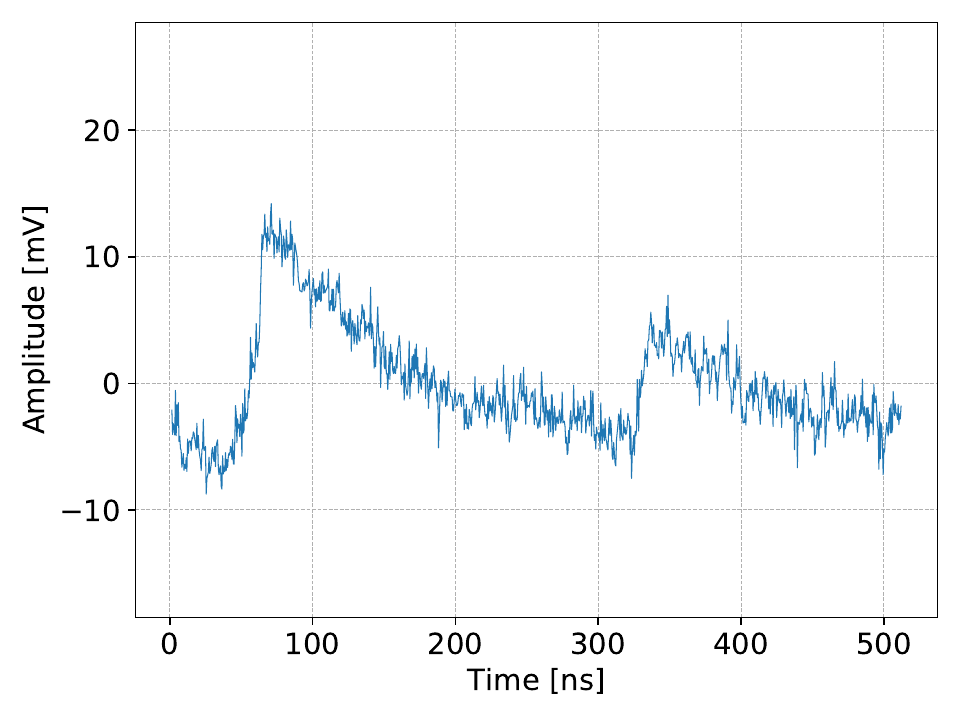}
\caption{Example of a signal trace. The sampling rate corresponds to 2\,Gsample/s. The displayed exampel was selected to represent a typical signal trace. The expectation value for the number of dark counts given the sensor temperature and the measured dark count rate is 3.5. }
\label{fig:trace}
\end{figure}

Dark counts are recorded triggering random readout events without ambient light to minimize the background of impinging photons. For each telescope, \num{100000}~events were triggered at a frequency of \SI{80}{\hertz}. The spatial temperature average of the sensors of both focal planes was \SI{28}{\degreeCelsius} with a typical spread of less than \SI{1}{\degreeCelsius}. All individual sensors were stable during data taking within \SI{0.2}{\degreeCelsius}. The channels are read out with the maximum sampling depth of \num{1024} samples per channel at a sampling rate of 2\,Gsample/s. An example of a signal trace is shown in Figure~\ref{fig:trace}.

A similar study on dark count spectra has been published by FACT~\cite{Biland_2014}, which uses the same data acquisition system and software as described in~\cite{Anderhub_2013}. Therefore, the same algorithm to extract the pulses from the recorded data was adopted. As it is explained in detail there, the approach is described here only briefly for completeness.

In the first stage, a baseline is determined and subtracted for each channel individually. As some channels suffer from periodic electronic noise, a sliding average with an adapted length of 10 samples, which corresponds to \SI{5}{\nano\second}, is applied. The resulting trace is scanned for a threshold crossing of \SI{5}{\mV} between two consecutive samples. Then, a local maximum is searched between 5 and 35~samples after the threshold crossing. The arrival time is defined by the last sample within the 30~samples before the local maximum with an amplitude that does not fall \SI{50}{\percent} of the maximum value. Integrating the raw signal for \SI{15}{\nano\second} (30~samples) starting at the arrival time yields the integrated charge. In addition to the original algorithm, the first turning point before the leading edge, i.\ e.\ the first sample with a larger amplitude than the preceding sample, is determined and subtracted as baseline from the raw signal before integration. An artificial dead time comparable to the total pulse length of \SI{100}{\nano\second} is introduced in the algorithm to skip the falling edge of detected pulses. Since the pulse extraction algorithm is designed to identify SiPM-like pulses, misidentification of electronics noise and the corresponding pedestal-peak in the spectrum are highly suppressed. 

Since the signals of three pixels in each camera are additionally split to provide an additional calibration output, their measured signal is attenuated by~\SI{18}{\percent} compared to the other 61~pixels. 

\subsection*{Spectrum Function}
The properties of the sensors are extracted from their dark count spectrum. To properly describe the measured dark count spectrum, one has to account for the probability $P_N$ to measure a multiplicity of $N$ breakdowns induced by the initial breakdown of a single cell as well as noise that smears out this distribution. In~\cite{Biland_2014} various distribution functions are compared. It is demonstrated that the best description of the dark count spectra is the modified Erlang distribution. In particular at high multiplicities, it outperforms other distribution functions, in particular the widely used Borel distribution. The coefficients of the modified Erlang distribution are given by
\begin{equation}
\label{eq:nu}
    P_N=\frac{(Nq)^{N-1}}{{[(N-1)!]}^\nu}\quad\textrm{with}\quad q\equiv p\cdot e^{-p}.
\end{equation}
The fit parameter $\nu$ originates from the fact that the number of potential crosstalk triggers by each breakdown is limited by the geometry of the sensor and is usually $\nu\sim1$. The parameter $p$ describes the probability for a breakdown to trigger an additional breakdown in a neighboring cell and should  not be confused with what is generally referred to as crosstalk probability. The crosstalk probability $p_\textrm{xt}$ can be derived as
\begin{equation}
p_\textrm{xt}=\frac{\smashoperator[r]{\sum_{n=2}^{\infty}}P_n}{\smashoperator[r]{\sum_{n=1}^{\infty}}P_n}.
\end{equation}
Two types of noise sources are considered in this analysis. The charge released in each avalanche undergoes cell-to-cell fluctuations and thus scales with the number of breakdowns. The electronics noise is independent of the multiplicity. Both types of noise are assumed to be Gaussian. Therefore, the measured dark count spectrum can be described by a sum of Gaussian distributions for multiplicity $N$ with width given by,
\begin{equation}
    \sigma_\mathrm{N}=\sqrt{\sigma_{el}^2+N\sigma_{pe}^2}
\end{equation}
where $\sigma_{el}$ and $\sqrt{N}\sigma_{pe}$ are the Gaussian widths for the electronics noise and the amplitude-dependent charge fluctuations respectively. Electronics noise in this context means noise of the whole electronics chain, including the kTC noise of the SiPM itself.
The resulting distribution can be written as
\begin{equation}
    \label{eqt:spectrum_function}
    f(x)=A_1 \, a_1 \, \cdot \sum_{n=1}^{n=\infty}P_n \frac{e^{-\frac{1}{2}[\frac{x-x_n}{\sigma_n}]^2}}{a_n}\ .
\end{equation}
The parameter $A_1$ denotes the amplitude of the single-p.e.\ peak. Introducing the gain $g$ and a baseline shift $x_0$, the position of the mean of each Gaussian and their normalization are defined as 
\begin{equation}
    x_n=x_0+n\cdot g\quad\textrm{and}\quad a_n=\sigma_n \sqrt{2\pi}\ .
\end{equation} 
Integration of $f(x)$ over the whole range of extracted signals results in the total number of dark counts and can be used to calculate the dark count rate if the extraction efficiency is known.\\

For each SiPM, the integrated signals are filled into a histogram. The resulting spectra of five SiPMs show unexpected features which are related to periodic noise in their traces. Two of these sensors are located at identical places in their corresponding camera. The location of the third sensor coincides with the missing sensor in the other. As the affected sensors have the same but indistinguishable locations in both cameras, the cause must be related to the electronics and can not be resolved remotely. Therefore, the five affected SiPMs are excluded from the analysis. 

The individual spectra are then fitted with a log-likelihood method with the distribution described in Equation~\ref{eqt:spectrum_function}. Since the fit result does not strongly depend on the exact value of the exponent $\nu$ (see Equation \ref{eq:nu}) and in a free fit, all fits yield results consistent with one, the exponent has been fixed to one in all fits. As expected, the obtained baseline offset $x_0$ is negligible compared to the gain, as the baseline has been subtracted during signal extraction.

\begin{figure}[ht]
\centering
\includegraphics[width=\textwidth]{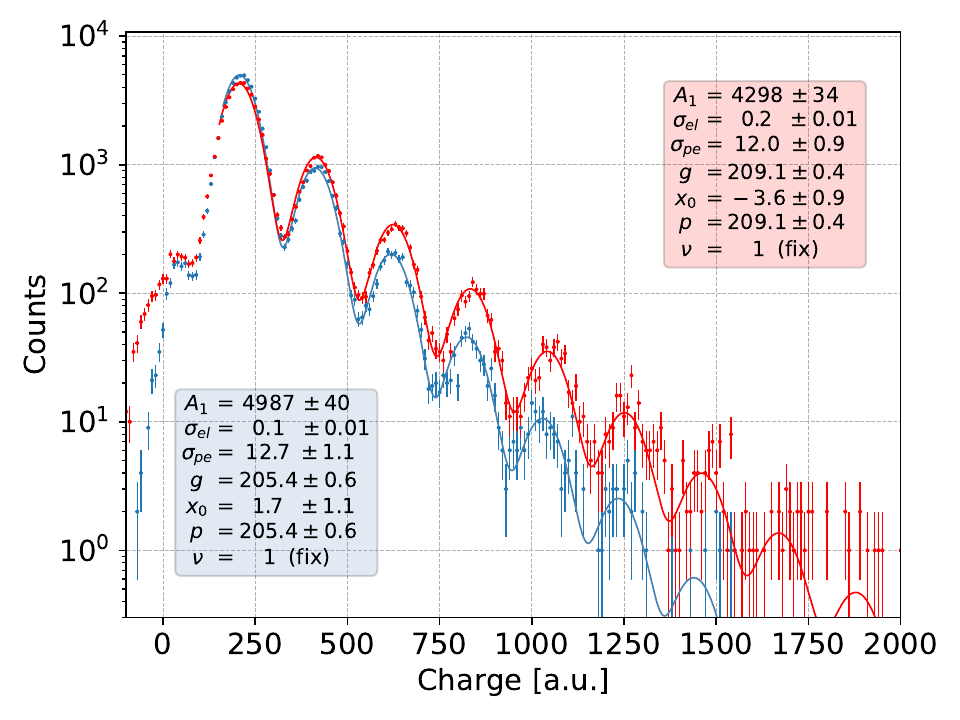}
\caption{Dark count spectrum extracted from a single SiPM optically coupled to a light guide (blue) and a blind pixel without a light guide (red). The lines show the fit to the distribution as introduced in Equation~\ref{eqt:spectrum_function} with the exponent $\nu$ fixed to the value extracted from the compiled spectrum shown in Figure~\ref{fig:spectrum}. All fit parameters are displayed in the colored boxes.}
\label{fig:spectrum_single}
\end{figure}

\begin{figure}[ht]
\centering
\includegraphics[width=\textwidth]{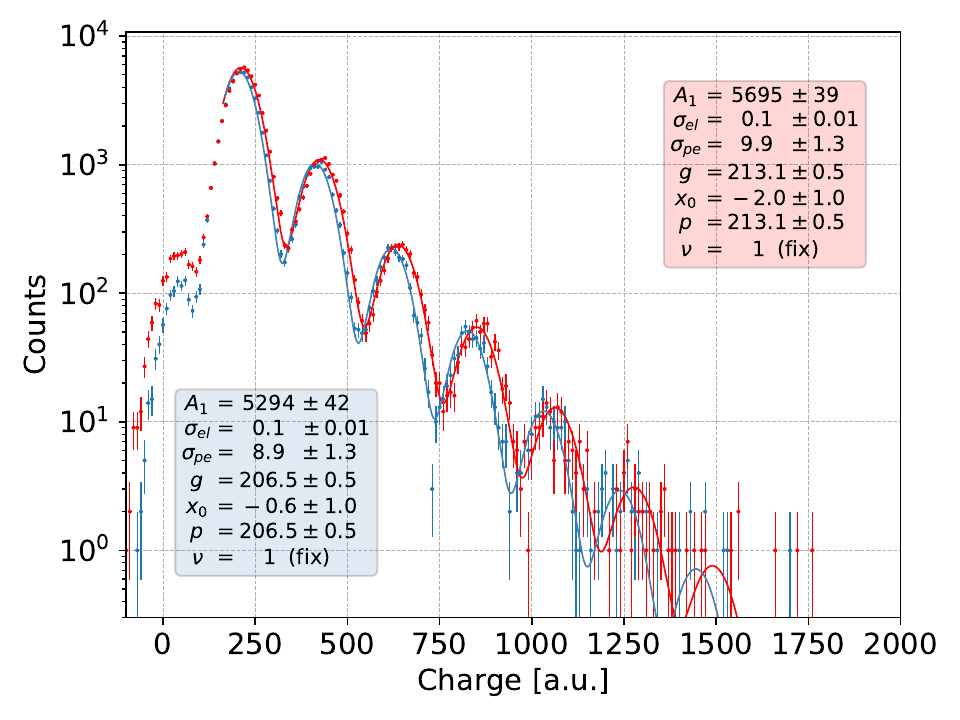}
\caption{Dark count spectra extracted from a two randomly chosen SiPMs coupled to a light guide. The lines show the fit to the distribution as introduced in Equation~\ref{eqt:spectrum_function} with the fit parameters displayed in the colored boxes.}
\label{fig:spectrum_single_comparison}
\end{figure}

An example of a fit to the spectrum of two single SiPMs is shown in Figure~\ref{fig:spectrum_single}. It compares the dark count spectrum of a SiPM optically coupled to a light guide to a blind pixel without a light guide. The fit is used to determine properties of each pixel individually, including the relative gain and the crosstalk probability. Figure~\ref{fig:spectrum_single_comparison} shows the dark count spectrum of two randomly chosen SiPMs with a light guide. It illustrates, that typical sensor-to-sensor fluctuations are considerably smaller than the difference between sensors with optically coupled light guides and bare sensors.

\begin{figure}[ht]
\centering
\includegraphics[width=\textwidth]{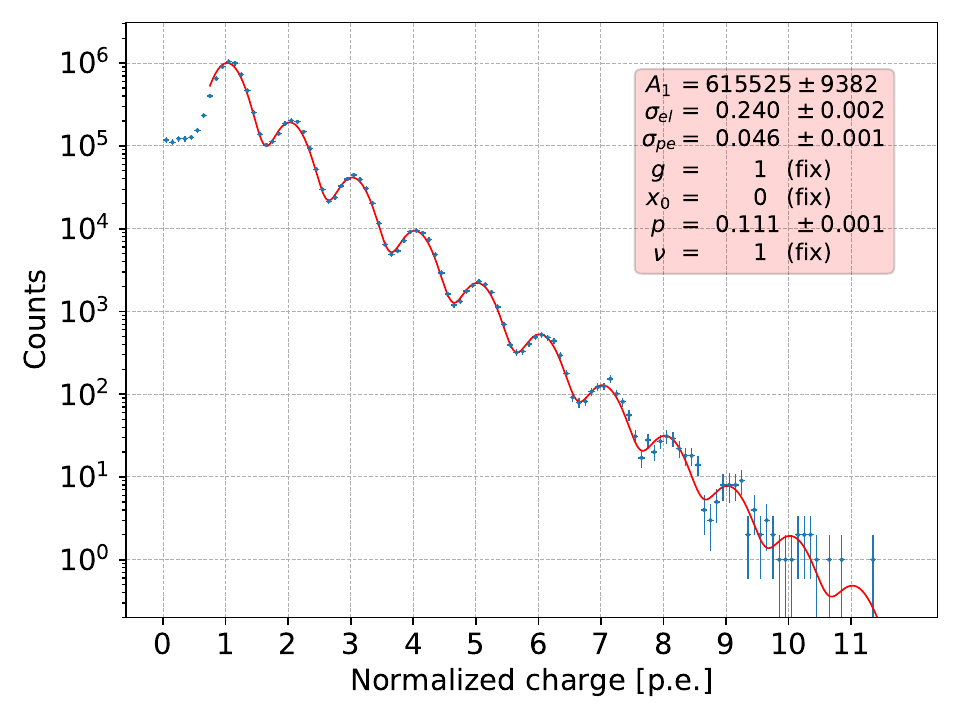}
\caption{Normalized dark count spectrum compiled from the majority of all SiPMs that are expected to have identical properties, i.~e.\ which are optically coupled to a light guide and do not suffer a reduced gain. The red line shows the fit as introduced in Equation~\ref{eqt:spectrum_function} with fixed values for the exponent ($\nu=1$) and the gain ($g=1$). 
}
\label{fig:spectrum}
\end{figure}

To compile all spectra into a single histogram, the baseline offset $x_0$ and the gain $g$ are used to normalize the individual signals to their multiplicity $N$. The resulting histogram is displayed in Figure~\ref{fig:spectrum}. To demonstrate proper normalization, in the displayed fit the baseline offset has been fixed to zero and the gain to one.

Indicating a measure of the fit quality is generally omitted as the purpose of the fit is not to verify the model but to extract reasonable parameters from the measured spectra, which is obviously achieved. All fits have been scanned manually and checked to represent the measured spectra reasonably well.

\section{Results} \label{sec:results}

All results follow from measurements at a temperature of \SI{28}{\celsius} and an overvoltage of \SI{5}{\volt}.

\begin{figure}[ht]
\centering
\hfill
\includegraphics[trim=3cm 0 3cm 0, clip,width=.480\textwidth]{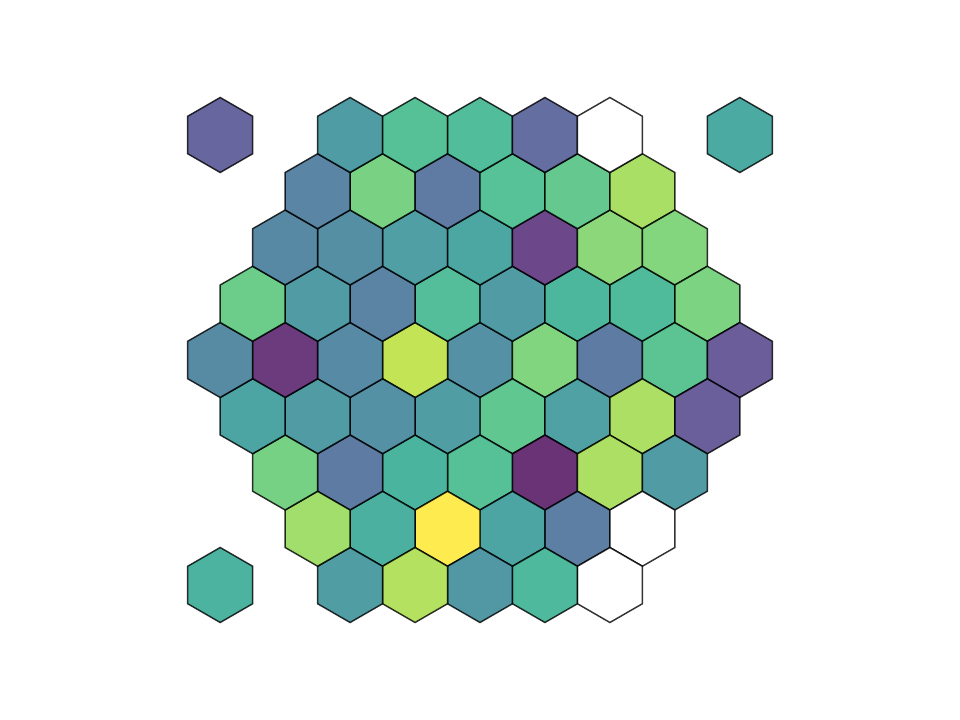}
\hfill
\includegraphics[trim=3cm 0 3cm 0, clip,width=.480\textwidth]{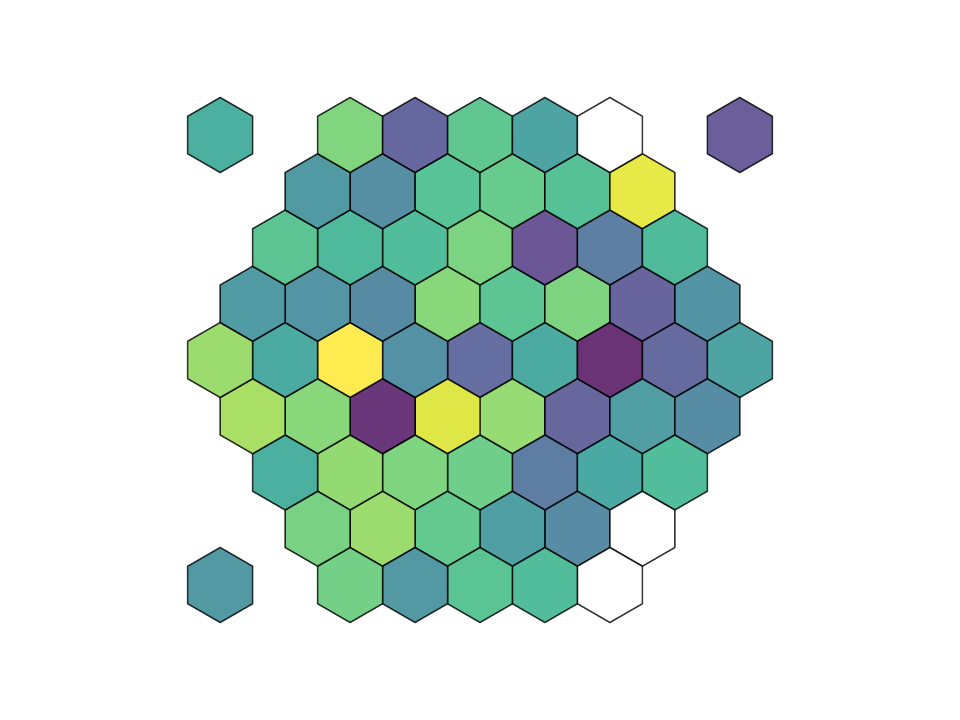}
\includegraphics[trim=0.2cm 0 0.2cm 0, clip,width=.96\textwidth]{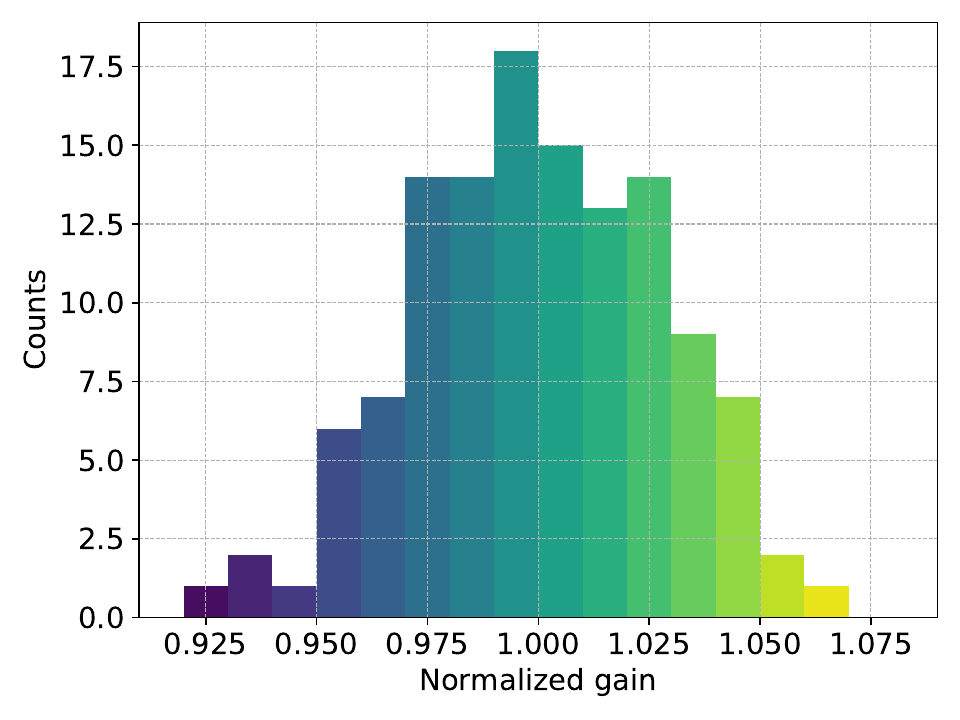}
\caption{Distribution of the resulting gain from a fit to the dark count spectrum of each of the 122~SiPMs normalized to the average gain. The gain of the six pixels with a lower electronics gain has been scaled accordingly. The arrangement in the two cameras shows a homogeneous distribution. The colors of the histogram bars match the colors of the camera display. The color scale ranges from purple (low) to yellow (high). Data taken at a temperature of \SI{28}{\celsius} and an overvoltage of \SI{5}{\volt}.}
\label{fig:gain}
\end{figure}

\paragraph{Gain Distribution.} Figure~\ref{fig:gain} shows the distribution of the gain normalized to the average gain of all 122~pixels included in the analysis. The distribution has a variance of roughly \SI{3}{\percent}. All values are within a range from \SI{92}{\percent} to \SI{107}{\percent}. The breakdown voltage has not been calibrated for each sensor individually. Thus, the applied bias voltage is set according to the data sheet values for the breakdown voltage and an intended overvoltage of \SI{5}{\volt}. The width of the distribution is well consistent with systematic deviations coming from the provided operation range of the manufacturer ($\pm\, \SI{5}{\percent}$) and the precision of the temperature feedback system~($\pm \,\SI{1}{\percent}$).

\paragraph{Dark Count Rate.} 

In general, the total number of extracted pulses allows determining the dark count rate of the sensors. However, this requires a precise knowledge of the efficiency of the method. For an overvoltage of \SI{5}{\volt} and a temperature of \SI{28}{\degreeCelsius} a dark count rate around \SI{200}{\kilo\hertz\per\square\milli\meter} is expected derived from the data sheet values. This corresponds to a total rate of over \SI{7}{\mega\hertz} for each sensor or roughly four pulses per trace. In contrast to the study in~\cite{Biland_2014}, the longer tails of the pulses produce significant pile-up reducing the efficiency of the described pulse extraction algorithm considerably. Without a detailed simulation, a precise estimate of the efficiency is not possible. Taking only the artificial dead time into account, the extracted pulse rate is around \SI{100}{\kilo\hertz\per\square\milli\meter}. However, statistically, about half of the pulses are expected on the falling edge (i.\ e.\ within \SI{100}{\nano\second}) of a preceding pulse and the pulse extraction is intentionally only sensitive to pulses starting close to the baseline. This extends the average dead time after a successfully detected pulse, i.\ e.\ the time of significantly reduced pulse extraction efficiency due to the tail of the pulse, by another \SI{50}{\percent} to typically at least \SI{150}{\nano\second}. Thus the measured number of pulses corresponds to at least \SI{150}{\kilo\hertz\per\square\milli\meter} which is reasonably close to the expectation. Additionally, the dark count rate of sensors of the same type can differ significantly for different wafers~\cite{LHAASO_2021}. Since all applied sensors come from the same wafer, one can not rule out a below-average dark count rate due to a slightly better production quality.

\begin{figure}[ht]
\centering
\hfill
\includegraphics[trim=3cm 0 3cm 0, clip,width=.48\textwidth]{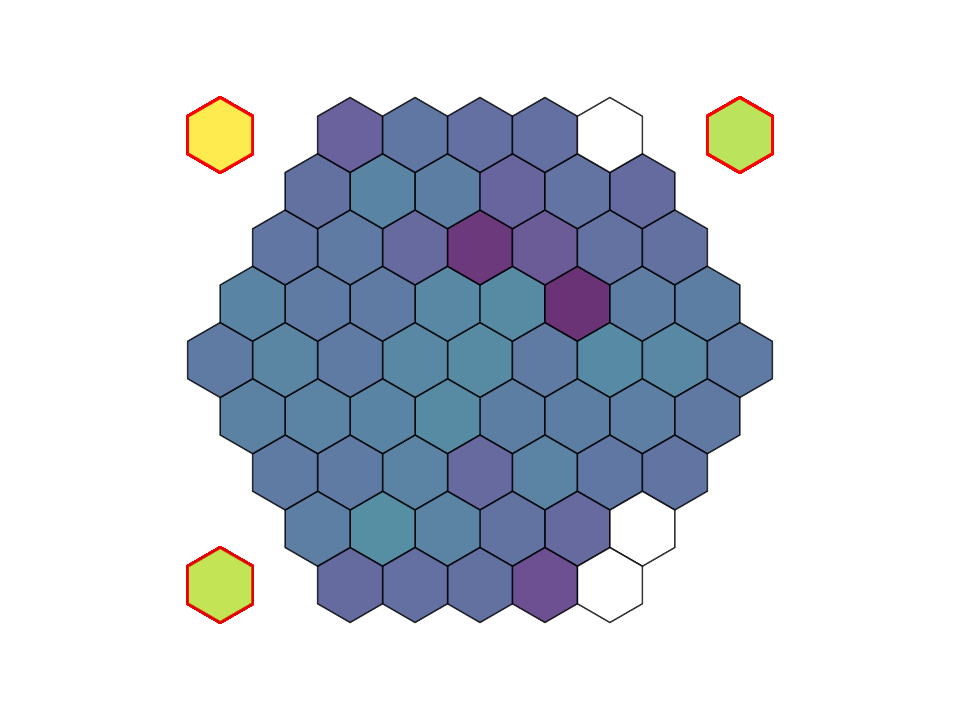}
\hfill
\includegraphics[trim=3cm 0 3cm 0, clip,width=.48\textwidth]{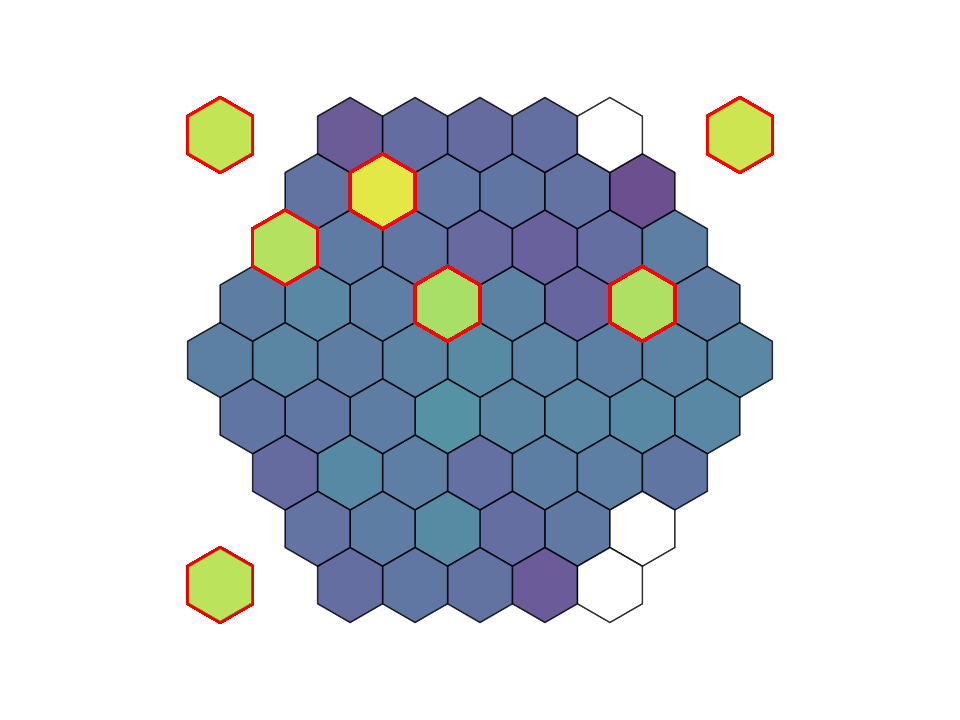}
\includegraphics[trim=0.2cm 0 0.2cm 0, clip,width=.92\textwidth]{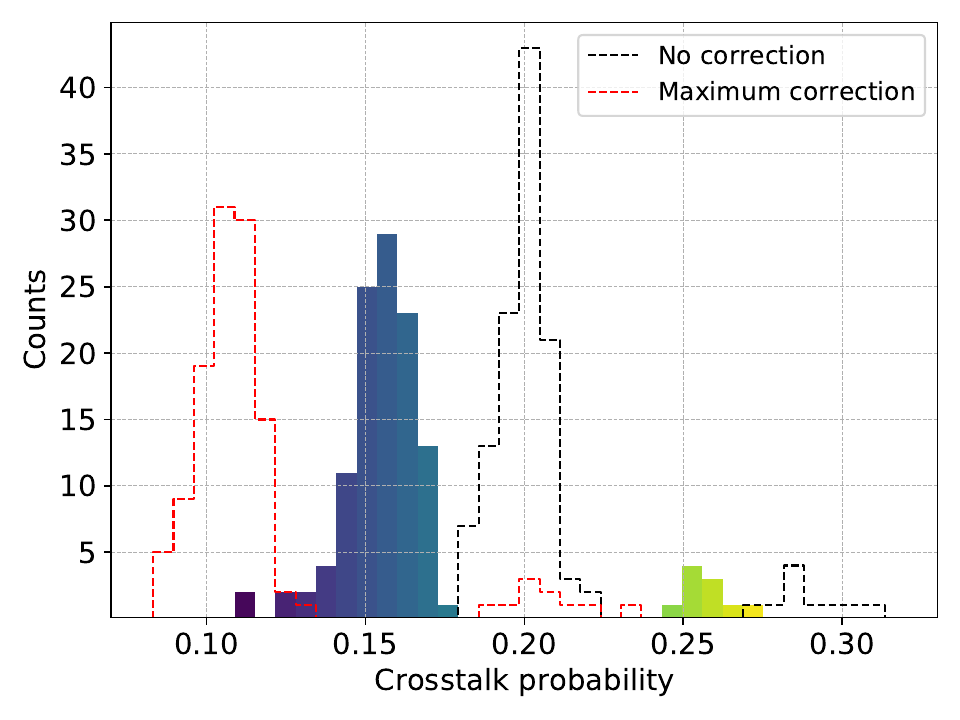}
\caption{Distribution of the resulting crosstalk probability from a fit to the dark count spectrum of 122~SiPMs accounting for coincident Poisson events. Two separated distributions are visible with a mean of $0.154 \pm 0.011$ and $0.257 \pm 0.012$ respectively. The locations of the corresponding pixels in the two cameras coincide exactly with the pixels with and without an optically coupled light guide. Pixels without coupling are marked in red. The colors of the histogram bars match the colors of the camera display. The color scale ranges from purple (low) to yellow (high). The dashed lines represent the same distribution without correction (black) and the maximum correction taking the dark count rate from the datasheet (red). Data taken at a temperature of \SI{28}{\celsius} and an overvoltage of \SI{5}{\volt}.}
\label{fig:crosstalk}
\end{figure}

\paragraph{Optical Crosstalk.} The distribution of the crosstalk probability $p_\textrm{xt}$ as evaluated by fitting the modified Erlang distribution to the measured dark count spectra is depicted in Figure~\ref{fig:crosstalk}. To account for coincident random breakdowns inside the integration window, the crosstalk probability is corrected as described in~\cite{Futlik_2011}. Using the measured dark count rate for each pixel, the crosstalk probability is corrected using formula (2) of the above paper. For completeness, the figure also shows the distribution without this correction as well as the correction using the datasheet value of \SI{200}{\kilo\hertz\per\square\milli\meter}. Two clearly distinct distributions are visible, corresponding to the SiPMs with and without an optically coupled light guide. As expected, the ten SiPMs without optically coupled light guides show a significantly larger crosstalk probability. The distribution of the coupled SiPMs has a mean of $0.154 \pm 0.011$, compared to the bare sensors with a mean crosstalk probability of $0.257 \pm 0.012$ corresponding to a relative reduction of \SI{40}{\percent}. The absolute crosstalk probability is larger than the value derived from the data sheet values which is 0.20 for an overvoltage of \SI{5}{\volt}. This value can be estimated by interpolating the data sheet values at \SI{2.5}{\volt} and \SI{6}{\volt} overvoltage assuming a linear dependence within this voltage range. Although such a linear dependence has been shown for other sensors~\cite{Dune_2016}, this linear interpolation has to be seen as an approximation and induces an uncertainty on the datasheet value of the crosstalk probability in the order of 0.01 to 0.02. Taking into account the uncertainties on the dark count rate, the datasheet value, and the width of the measured distribution, the measured crosstalk probability for the sensors without optical coupling is consistent with the expectation.
Many standard methods utilize rate measurements with fixed thresholds (e.\ g.\ at 0.5\ p.\ e.\ and 1.5\ p.\ e.) to determine a crosstalk probability. In this approach, identical pulses originating from different multiplicities remain indistinguishable, while the integration of the individual Gaussians in the distribution function correctly assigns them proportionally. The systematic difference has been checked by integrating the whole distribution function starting at 0.5\ p.\,e.\ and 1.5\ p.\,e.\ respectively, yielding a value higher by +0.02.

\section{Summary and Conclusions}
In this study, the characteristics of 122~SiPM sensors of type SensL MicroFJ-60030-TSV have been evaluated. This has been done using the cameras of two HAWC's Eye telescopes, applying readout electronics originally produced for the FACT telescope. Out of these 122~sensors, 112~sensors are properly coupled to solid optical light guides. The bias power has been provided by a power supply achieving a precision of \SI{\ll\,1}{\percent} that was neglected in this study. The temperature-based overvoltage stabilization works with a precision corresponding to \SI{1}{\percent} of the sensors' overvoltage, including systematic effects from uncalibrated temperature sensors. For temperature-based voltage compensation, the temperature coefficient from the data sheet was applied. The maximum range of allowed breakdown voltages given by the manufacturer corresponds to \SI{\pm\,5}{\percent} for an overvoltage of \SI{5}{\volt}. Therefore, the systematic contributions are dominated by the knowledge of the breakdown voltage.\\

The measurement of the relative gain of 122~sensors showed a maximum deviation from the mean value of \SI{\pm\,7.5}{\percent} with a variance of only \SI{3}{\percent}. This is well consistent with the data sheet of the manufacturer. While in a similar study of the FACT camera, the breakdown voltage for each sensor had still to be calibrated to achieve a variance of $2.4\,\%$, this study intentionally has omitted all calibrations related to the SiPMs themselves and applied a temperature-based voltage compensation circuit without additional absolute calibration of the temperature sensors.\\

In addition, the crosstalk probability of 112~sensors with, and nine without, optical coupling to light guides has been investigated. Since with optically coupled light guides, crosstalk photons that are reflected on the protective window can leave the SiPM, the probability for optical crosstalk has decreased. A reduction by \SI{40}{\percent} from an average probability of \SI{26}{\percent} for the bare sensors to an average probability of \SI{15}{\percent} has been confirmed for sensors with an optically coupled light guide. An additional single light guide shows an unclear status of optical coupling. As its crosstalk value falls exactly into the distribution of the nine pixels without light guide, it can be concluded that no proper coupling exists for this sensor as well.\\

The absolute values of the measured crosstalk probability and its reduction due to optical coupling to solid light guides can not be transferred to other SiPM types, as the amount of crosstalk photons that are reflected at the entrance window highly depends on the physical properties of the sensor, but it provides a good indication of the order of the effect.\\ 

Reduction of crosstalk probability was also investigated in a laboratory setup for the solid light guides used in the FACT telescope while they were coupled to SiPMs of type Hamamatsu S14520-3050\,VN. A relative decrease of the optical crosstalk probability by about \SI{29}{\percent}~\cite{Priv_Comm_Tajima} was observed. This result confirms the significant reduction demonstrated in this paper. As a different sensor type and light guides of different geometrical properties were applied, the measured values can not be compared directly.\\

The analysis of the performance of the photo sensors installed in the two HAWC's Eye telescopes confirms the high precision of recent SiPMs. It demonstrates that even for high precision requirements, the application of data sheet values is sufficient. Apart from a reasonable precise power supply, no additional calibration is necessary. This study also confirms that coupling the sensors to well-designed light guides reduces optical crosstalk between neighboring cells significantly.

\clearpage
\section*{Acknowledgments}
We thank Johannes Schumacher for his excellent design of the bias power supply, and the electronics and mechanics workshop of the Physics Institute 3A and 3B of RWTH Aachen University for their support. We also acknowledge Kazuhiro Furuta, Akira Okumura, and Hiroyasu Tajima for their measurement cited in the conclusions. This work was partially funded by the Excellence Initiative of the German federal and state governments and the German Academic Exchange Service, the DGAPA PAPIIT (project number IG101320) and the CONACYT-DAAD (project 279446).

\newcommand{\arXiv}[1]{{\href{http://arxiv.org/abs/#1}{arXiv:#1}}}

\end{document}